\documentclass[16pt]{article}
\usepackage{graphicx,amsfonts,latexsym}
\usepackage{algorithmic}
\usepackage{algorithm}
\usepackage{amsmath,amsfonts,amssymb,amsmath}

\newtheorem{theorem}{Theorem}
\newtheorem{lemma}{Lemma}
\newtheorem{proposition}{Proposition}
\newcommand{\reff}[1]{(\ref{#1})}

\usepackage[cp1251]{inputenc}

\newcommand{\y}{{{\bf y}}}

\newcommand{\wh}[1]{\widehat{#1}}

\newcommand{\ola}[1]{\overline{\alpha}}

\def\bbP{\mathbb P}
\def\bbR{\mathbb R}
\def\bbZ{\mathbb Z}
\def\bfm{\boldsymbol}

\def\diy{\displaystyle}
\def\dist{{\rm{dist}}}

\begin{document}
\title{\bf A non-linear model of trading mechanism on a financial
market}

\author{N.Vvedenskaya$^1$ ,
Y.Suhov$^{1,2,3}$, V.Belitsky$^3$}

\date{}
\maketitle


\begin{abstract} We introduce a prototype model in an attempt to capture some
aspects of market dynamics simulating a trading mechanism. The model
description starts with a discrete-space, contiinuous-time Markov
process describing arrival and movement of orders with different
prices. We then perform a re-scaling procedure leading to a
deterministic dynamical system controlled by non-linear odinary
differential equations (ODEs). This allows us to introduce
approximations for the equilibrium distribution of the model
represented by fixed points of deterministic dynamics. \vskip 5
truemm
\end{abstract}

\section{Introduction}

This paper proposes a model that takes into account, in a rather
stylized form, some aspects of automated trading mechanisms adopted in
modern financial markets, in particular, the dynamics of the limit
order book.

In short, a limit order book keeps records of arivals, movements and
departures of market participants (traders) who declare their trading positions.
An arriving trader may wish to buy or sell at a certain price, and can move
his declared price when time progresses. If the declared price is met by
a trader with the opposite intention, a trade is recorded: this may lead to
disappearance of one or both participants from the market, due to exhaustion
of their offers.
For a detailed description of some common limit order book models
and their applications, see \cite{CST,CKS,R} and references therein.

-----------------------------------------------

$^1${Institute for Information Transmission Problems, RAS, Moscow, Russia}

$^2${DPMMS, University of Cambridge, and St John's College, Cambridge, UK}

$^3${IME, Universidade de Sao Paulo, Sao Paulo, Brazil}
\vfill\eject

In the current paper we present a somewhat different model, including
elements of
queueing behavior of arrived offers; this model is studied by using
techniques of asymptotic analysis. An earlier account of this work
(in its preliminary version) can be found in \cite{VSB}. Compared
with \cite{VSB}, in the present text we adopt a continuous-time
setting for the basic Markov process: it clarifies the meaning of
the main parameters of the model and shortens the proof of some of
our main results.

The model under consideration is a prototype; at this stage it does
not aim to take
into account all possible aspects that can be viewed as defining,
either theoretically or practically. Instead, we opted for a
simplified description which leads to some straightforward, yet
instructive, answers.

Our model differs from known models of the limit order book in a
number of aspects. Arguably, it can be a subject of criticism (as a
number of other proposed models). In particular, the strategic
behavior of the model in its current form (and some further details)
do not quite match existing mechanisms governing electronic trading
on financial markets. Nevertheless, the model shows a certain amount
of flexibility, and covers a borad range of situations. Its mathematical
advantage is that in the scaling limit under consideration, it leads to
a single fixed point.

Our scaling limit is based on suppositions that (i) the number of market
participants
is large, (ii) during a very short time period only part of them makes a
decision of performing a trade or maiking a move along the price range,
and (iii) the probability for any given participant to make such decision
is small. This makes it natural to change, in a suitable
manner, parameters of original Markov process.

After rescaling, a limiting dynamical system emerges, with a deterministic
behavior described by a system of non-linear ordinary differential equations.
The rescaling techniques greatly simplify the structure of the model,
and this phenomenon extends far beyond basic examples like
the current prototype model. As we mentioned earlier, the present paper
focuses on a simplified model, with `minimal' number of constant
parameters, where some of technically involved issues are absent.

A similar approach is commonly used in the literature on stochastic
communication networks; see, e.g., \cite{VDK,M,MS,VS} and
\cite{SGK}. We also find similarities, as well as differences, with
models proposed (in a different context) in a recent paper
\cite{MMZ}; analogies with \cite{MMZ} could be useful for the
aforementioned purpose of defining the prices that are appropriate
for trades.

In the next section we describe the underlying Markov process. In
Section 3 the rescaling of the process is presented and the main
results are stated and the proofs  are given. The last section
contains concluding discussion of various aspects of the model.
\medskip

\section{ The underlying Markov process}

The rationale for the models below is as follows. We consider a
single-commodity market where prices may be at one of $N$ distinct
levels (say, $c_1<c_2<\ldots <c_N$, although the exact meaning of
these values is of no importance here).

The market is operating in continuous time $t\in\bbR_+$ where
$\bbR_+=[0,\infty)$. (As was mentioned above, the earlier version
\cite{VSB} used a more cumbersome discrtete-time version of the
underlying process.) At a given time $t\in\bbR_+$, there are
$b_i(t)$ traders prepared to buy a unit of the commodity at price
$c_i$ and $s_i(t)$ traders prepared to sell it at this price, which
leads to vectors

\begin{equation*}
\bfm{b}(t)=\big(b_1(t),\ldots ,b_N(t)\big),\;\;
\bfm{s}(t)=\big(s_1(t),\ldots ,s_N(t)\big)\in\bbZ_+^N.
\end{equation*}
Here and below $\bbZ_+=\{0,1,\ldots\}$ stands for a non-negative integer
half-lattice and $\bbZ_+^N$ for the non-negative integer $N$-dimensional
lattice orthant. The pair $(\bfm{b}(t), \bfm{s}(t))$ represents a
state of a Markov process $\{\bfm{U}(t)\}$ that will be the subject of
our analysis. \def\RT{\rm T} \def\RQ{\rm Q} \def\RM{\rm M} \def\RB{\rm B}
\def\RS{\rm S} \def\RB{\rm B} \def\RS{\rm S}\def\oo{\overline o}

Suppose that, for given $k=1,\ldots N$ and $t\in\bbR_+$, we have that
$b_k(t)\geq s_k(t)>0$ then each of the sellers gets a trade at a given
rate $\rho_{\,\RT}>0$ and leaves the market, together with one of the buyers.
Therefore, both values $b_k(t)$ and $s_k(t)$ decrease at rate $\rho_{\,\RT}$.
In addition, each one among $s_k(t)$ sellers (i) quits the
market at rate $\rho_{\,\RQ}>0$ or (ii) moves to the price level $c_{k-1}$
at rate $\rho_{\,\RM}>0$, if $k>1$. Similarly, every buyer among the $b_k(t)$
buyers (i) quits the market at the same rate $\rho_{\,\RQ}$ as above or (ii)
moves to the price level $c_{k+1}$, again at rate $\rho_{\,\RM}$, provided
that $k<N$.

Symmetrically, if $s_k(t)\geq b_k(t)>0$ then each of the buyers gets
a trade at rate $\rho_{\,\RT}$ and leaves the market,
together with his seller companion. The remaining traders at the
price level $c_k$ proceed in a manner as above.

Further, when $k=N$, a buyer leaves the
system with rate $\rho_{\,\RQ}+\rho_{\,\RM}$.
Similarly, for $k=1$, a
seller leaves the system with rate $\rho_{\,\RQ} + \rho_{\,\RM}$.

Finally, a random Poisson flow of new exogenous buyers arrives at
the price level $c_1$; the rate of this arrival equals
$\lambda_{\RB}>0$. Similarly, a Poisson random flow of new sellers
arrives at the price level $c_N$; the rate of this arrival is
$\lambda_{\RS}>0$.

As usually, standard independence assumptions are in place.
This generates the aforementioned Markov
process $\big\{\bfm{U}(t)\big\}$ with trajectories
$\big\{(\bfm{b}(t),\bfm{s}(t))\big\}$, $t\in\bbR_+$.

\medskip

\begin{theorem}\label{thm1}
For any values of parameters $\lambda_{{\RB}/ {\RS}}$,
$\rho_{\,{\RQ}/{\RM}}$ and $\rho_{\,\RT}$, the process
$\{\bfm{U}(t)\}$ is irreducible, aperiodic and positive recurrent.
Therefore, it has a unique set of equilibrium probabilities
$\pi=\Big(\pi\big(\bfm{b},\bfm{s}\big):\;\bfm{b},\bfm{s}\in\bbZ^N_+\Big)$,
and
for any initial state $\bfm{U}(0)$ (deterministic or random), the
distribution of the random state $\bfm{U}(t)$ at time $t$ converges
weakly to $\pi$ as $t\to\infty$:
\begin{equation*}
\lim_{t\to\infty}\bbP\Big(\bfm{U}(t)=(\bfm{b},\bfm{s})\Big)
=\pi\big(\bfm{b},\bfm{s}\big).
\end{equation*}
\end{theorem}

\medskip

{\it Proof of Theorem} \ref{thm1}. Irreducibility and aperiodicity of the
process is evident. Positive recurrence follows from the
following observation. The (random) time that a given trader (a
buyer or a seller) spends in the system, i.e., the time from his
arrival till exit, is majorized by a sum of $N$ independent
exponential variables. Therefore, the process $\{\bfm{U}(t)\}$ can
be majorized, in a natural fashion, by an M/M/$\infty$ queueing
process. But the latter is known to be positive recurrent.

The remaining assertions of Theorem \ref{thm1} are standard.
$\quad\blacktriangle$ 

\medskip

Despite a concise description, the detailed pattern of behavior of
process $\{\bfm{U}(t)\}$ is rather complex, particularly for
large values of $N$. For instance, consider the differences between
vectors $\bfm{b}(t')$ and $\bfm{b}(t)$ and between
$\bfm{s}(t')$ and $\bfm{s}(t)$, on a time interval $(t,t')$ where $0<t<t'$.
The increments for the entries $b_k(\,\cdot\,)$ and $s_k(\,\cdot\,)$ for
$1\leq k\leq N-1$ are captured by the following equations:
$$b_k(t')=b_k(t)+i_{k-1}^{\,\RM}(t,t')-n_{k}(t,t')-i_{k}^{\,\RM}(t,t')-i_{k}^{\,\RQ}(t,t'),$$
and
$$s_k(t')=s_k(t)+j_{k+1}^{\,\RM}(t,t')-n_{k}(t,t')-j_{k}^{\,\RM}(t,t')-j_{k}^{\,\RQ}(t,t').$$
Here $i_{k}^{\,\RM}(t,t')$ is the number of buyers who move within
time interval $(t,t')$ from level
$k$ to $k+1$ and $j_{k}^{\,\RM}(t,t')$ that of sellers who  move from level
$k$ to $k-1$. Next, $i_{k}^{\,\RQ}(t,t')$ is the number of buyers who quit the system
during interval $(t,t')$
from level $k$ and $j_{k}^{\,\RQ}(t,t')$ the number of sellers who quit the system
from level $k$. Finally, $n_k(t,t')$ is the number of buyers and sellers who got a
trade over $(t,t')$ at level $k$. All listed quantities are non-negative integer-values
random variables. For $k=1$ the structure of the expression is
similar, with the term $i_{k-1}^{\,\RM}(t,t')$ being replaced by $i_{\,\RB}(t,t')\geq 0$,
while for $k=N$ the term $j_{k+1}^{\,\RM}(t,t')$ is replaced by
$j_{\,\RS}(t,t')\geq 0$; both $i_{\,\RB}(t,t')$ and $j_{\,\RS}(t,t')$ being distributed
according to a Poisson law with mean $\lambda_{\RB / \RS}(t'-t)$.

In the simplest case of a market with one price level ($N=1$), the
process $\{\bfm{U}(t)\}$ is a continuous-time random walk on the
two-dimensional lattice quadrant $\bbZ_+^2$, where
$$b(t')=b(t)+i_{\,\RB}(t,t')-n(t,t')-i_{\,\RQ}(t,t')$$
and
$$s(t')
=s(t)+j_{\,\RS}(t,t')-n(t,t')-j_{\RQ}(t,t').$$
This already makes analytical representations for the
invariant distribution $\pi$ rather complicated; cf.
\cite{KS} and references therein.

\section{ Scaling limit}

The complexity of the time-dynamics and of the equilibrium distribution $\pi$
for process
$\big\{\bfm{U}(t)\big\}$ makes it desirable to develop efficient
methods of approximation. In this paper we focus on one such method
based on scaling the parameters of the process (including states and
time-steps).

The re-scaling procedure is as follows: we fix values
$\gamma >0$, $\beta >0$, $\alpha >0$,
$\lambda_{\RB}>0$ and $\lambda_{\RS}>0$ and set:
\begin{equation}\label{3}
\rho_{\,\RT}=\frac{\gamma}{L},\;\;\;\rho_{\,\RQ}
=\frac{\beta }{L},\;\;\;\rho_{\,\RM}
=\frac{\alpha }{L}.
\end{equation}

In addition, we re-scale the states and the time: pictorially,
\begin{equation}\label{3A}x_k\sim \frac{b_k}{L},\;\;y_k\sim \frac{s_k}{L},\;\;
\tau\sim\frac{t}{L}.
\end{equation}
Formally, denoting the Markov process generated for a given $L$
by $\bfm{U}^{(L)}$, we consider the continuous-time process
\begin{equation}
\label{4}
\bfm{V}^{(L)}(\tau )={\diy\frac{1}{L}}\bfm{U}^{(L)} \big(\tau
L ),\;\;\tau\geq 0.
\end{equation}

Let $\bbR_+^N$ denote
a positive orthant in $N$ dimensions. Suppose we are given a pair of
vectors $(\bfm{x}(0),\bfm{y}(0))\in\bbR_+^N\times\bbR_+^N$ where
$\bfm{x}(0)=(x_1(0),\ldots ,x_N(0))$, $\bfm{y}(0)=(y_1(0),\ldots ,
y_N(0))$. Consider the following system of first-order ODEs for
functions $x_k=x_k(\tau )$ and $y_k=y_k(\tau )$ where $\tau >0$ and
$1\leq i\leq N$ :
\begin{equation}\label{5}
\begin{array}{cclc}
{\dot x}_1&=&\lambda_{\RB}-\Big(\beta +\alpha \Big)x_1
 -\gamma\min\big[x_1,y_1\big],&\\
{\dot x}_k&=&\alpha x_{k-1} -\Big(\beta +\alpha \Big)x_k
-\gamma\min\big[x_k,y_k\big],& 1<k\leq N,\\
{\dot y}_k&=&\alpha y_{k+1} -\Big(\beta +\alpha \Big)y_k
 -\gamma\min\big[x_k,y_k\big],& 1\leq k< N,\\
{\dot y}_N&=&\lambda_{\RS}-\Big(\beta + \alpha \Big)y_N
 -\gamma\min\big[x_N,y_N\big],&\end{array}
\end{equation}
with the initial data
\begin{equation}\label{d}
 x_k(0)\geq 0, \ y_k(0)\geq 0, \ \ 1\leq k\leq N.
\end{equation}
The fixed point $\big(\bfm{x}^*,\bfm{y}^*\big)$ of system (\ref{5}) has
$$\bfm{x}^*=(x^*_1,\ldots ,x^*_N)\;\hbox{ and }\;\bfm{y}^*=(y^*_1,\ldots
,y^*_N)$$ 
where $x^*_k$ and $y^*_k$ give a solution to
\begin{equation}\label{6}
\begin{array}{cclcc}
\lambda_{\RB}&=&\Big(\beta +\alpha \Big)x^*_1
+\gamma\min\;\big[x^*_1,y^*_1\big],& &^{(1')}\\
\alpha x^*_{k-1}&=&\Big(\beta +\alpha \Big)x^*_k
+\gamma\min\;\big[x^*_k,y^*_k\big],&1<k\leq N,&^{(2')}\\
\alpha y^*_{k+1}&=&\Big(\beta +\alpha \Big)y^*_k
+\;\gamma\min\;\big[x^*_k,y^*_k\big],&1\leq k< N,&^{(3')}\\
\lambda_{\RS}&=&\Big(\beta + \alpha  \Big)y^*_N
+\gamma\min\;\big[x^*_N,y^*_N\big].& &^{(4')}\end{array}
\end{equation}
(In  \reff{6} we noted individual equations by addition signs that
are used below). Both systems (\ref{5}) and (\ref{6}) are
non-linear. However, the non-linearity `disappears' at a local level
which greatly simplifies the analysis of these systems.

In Theorems 2 and 3 below, we use the distance generated by the
Euclidean norm in $\bbR^N\times\bbR^N$.

\medskip

\begin{theorem}\label{thm2}
 {\sl {\rm{(a)}}} For
any  initial data $(\bfm{x}(0),\bfm{y}(0))$ with $x_i(0)\geq 0,\
y_i(0)\geq 0$, $1\leq i\leq N$ for all $\tau>0$ there exists a
unique  solution $(\bfm{x}(\tau ),\bfm{y}(\tau ))$, to problem
{\rm{\reff{5}--\reff{d}}} and $x_i(\tau)>0,y_i(\tau)>0$.

{\rm{(b)}} As
$\tau\to\infty$, the solution approaches a fixed point which is a
unique solution to system {\rm{\reff{6}}}:
\begin{equation}\label{7}
\dist\Big[\big(\bfm{x}(\tau ),\bfm{y}(\tau )\big),
\big(\bfm{x}^*,\bfm{y}^*\big)\Big]\to 0.
 \end{equation}
\end{theorem}

\medskip

{\it Proof of Theorem} \ref{thm2}, (a). Obviously, a unique solution
exists for sufficiently small $\tau >0$. Suppose that $
x_i={\operatornamewithlimits{\max}\limits_k}\;x_k $ and
${\operatornamewithlimits{\max}\limits_{k<i}}\;x_k<x_i$. Then
$\dot{x}_i\leq 0$. Similarly if  $y_j={\operatornamewithlimits{
\max}\limits_k}\;y_k$ and
${\operatornamewithlimits{\max}\limits_{k>j}}\;y_k<y_j$ then
$\dot{y}_j\leq 0$. Also, $\dot{x}_1<0$ if $x_1>\lambda_{\RB}/(\beta
+\alpha )$ and $\dot{y}_N<0$ if $y_N>\lambda_{\RS}/(\beta +\alpha
)$. Moreover, $\dot{x}_1\geq 0$ when $x_1=0$ and $\dot{y}_N\geq 0$
when $y_N=0.$ Thus the components of the solution $x_i (\tau )$ and
$y_i (\tau )$ are non-negative and uniformly bounded. By standard
constructions of the ODE theory, a unique solution
$\Big\{(\bfm{x}(\tau ),\bfm{y}(\tau ))\Big\}$ exists for all
$\tau>0$, and $\bfm{x}(\tau ),\bfm{y}(\tau )\in\bbR_+^N$.
$\quad\blacktriangle$

\medskip

Before proving assertion (b), we discuss several properties of the
solution to \reff{5}--\reff{d}. The following Proposition
\ref{prop1} indicates that the solution to \reff{5} possesses a kind
of the min/max principle. \vskip 5 truemm

\begin{proposition}\label{prop1}
Suppose that we have two initial points,
$\bfm{x}=(x_1,\ldots ,x_n)$ and $\bfm{x}'=(x'_1,\ldots ,x'_n)\in\bbR_+^N$
such that for some $k=1,\ldots ,N$,
\begin{equation}\label{d0}
x_k\leq x_k',\;\hbox{ and }\;y_k\geq y_k'.
\end{equation}
Then for the solutions $(\bfm{x}(\tau),\bfm{y}(\tau) )$,
$(\bfm{x}'(\tau),\bfm{y}'(\tau) )$, to {\rm{\reff{5}}}, with the
initial conditions $(\bfm{x}(0), \bfm{y}(0))=(\bfm{x},\bfm{y})$ and
$(\bfm{x}'(0), \bfm{y}'(0))=(\bfm{x}',\bfm{y}')$, for all $\tau
>0$,
\begin{equation}\label{ineq}
x_k(\tau)\leq x_k'(\tau),\;\hbox{ and }\;y_k(\tau)\geq y_k'(\tau).
\end{equation}
Similarly if $x_k\geq x_k'$ and $y_k\leq y_k'$ then the corresponding
solutions obey $x_k(\tau )\geq x_k'(\tau )$ and $y_k(\tau )\leq y_k'(\tau )$,
for all $\tau>0$
\end{proposition}

\medskip

{\it Proof of Proposition} \ref{prop1}. It is sufficient to consider
the situation where the strict inequalities take place. Suppose that
\reff{ineq} holds strictly for $\tau<\tau_0$ and fails at
$\tau=\tau_0$. For instance, let $x_k(\tau_0)=x_k'(\tau_0)$ and
assume that $k>1$ is the minimal index for which such an equality
takes place.

If $y_k(\tau_0)>y'_k(\tau_0)$ then
$\dot{x}_k(\tau_0)<\dot{x}'_k(\tau_0)$ and the above equality
$x_k(\tau_0)=x_k'(\tau_0)$
is impossible. The other case is considered in a similar manner.
$\quad\blacktriangle$ 
\medskip

A corollary of proposition \ref{prop1} is

\medskip

\begin{proposition}\label{prop2}
If $\dot{x}_k(0)\geq 0$ and $\dot{y}_k(0)\leq 0$ for all
$1\leq k\leq N$ then $x_k(\tau)$ increases and $y_k(\tau)$ decreases in
$\tau$, for all $\tau>0$ and $1\leq k\leq N$. Similarly if $\dot{x}_k(0)\leq 0$
and $\dot{y}_k(0)\geq 0$, $1\leq k\leq N$, then $x_k(\tau )$ decreases and
$y_k(\tau )$ increases in $\tau$.\end{proposition}

\medskip

{\it Proof of Proposition} \ref{prop2}. It again suffices to assume
that the strict inequalities hold true: $\dot{x}_k(0)> 0$ and
$\dot{y}_k(0)<0$. Then, for a small $\delta >0$:
$x_k(\delta)>x_k(0)$ and $y_k(\delta)<y_k(0)$.

Set $x'_k(0)=x_k(\delta),\ y'_k(0)=y_k(\delta)$. The coefficients of
equations do not depend on $\tau$, therefore $x'_k(\tau)=x_k(\tau
+\delta)$ and $y'_k(\tau)=y_k(\tau+\delta)$ for all $\tau >0$. By
Proposition 1, $x_k(\tau+\delta)=x'_k(\tau)\geq x_k(\tau)$ and
$y'_k(\tau+\delta)=y'_k(\tau)\leq y_k(\tau)$, for all $\tau>0$ and
$1\leq k\leq N$. As $\delta$ may be arbitrarily small, the assertion
of Proposition \ref{prop2} is valid. $\quad\blacktriangle$

\medskip

{\it Proof of Theorem 2}, (b). Given $\bfm{x}(0)=(x_1(0),\ldots
,x_N(0))$ and\\ $\bfm{y}(0)=(y_1(0),\ldots ,y_N(0))\in\bbR_+^N$, let
$(\bfm{x}(\tau ), \bfm{x}(\tau ))$ be the solution to (\ref{5}),
(\ref{d}). Consider two additional solutions,
$(\bfm{x}'(\tau),\bfm{y}'(\tau) )$ and $(\bfm{x}'' (\tau),\bfm{y}''
(\tau) )$, to (\ref{5}) with
$$x'_k(0)=0, \; y'_k(0)=\max \Big[\lambda_{\RS}/(\alpha +\beta ),\;\max_i
y_i\Big],$$
and
$$ x'' _k(0)=\max \Big[\lambda_{\RB}/(\alpha +\beta),\;\max_i
x_i(0)\Big], \; y'' _k(0)=0.$$

By Propositions \ref{prop1} and \ref{prop2}
\begin{equation}\label{xbod}
x'_k(\tau)\leq x_k(\tau)\leq x'' _k(\tau),\;\; y'_k(\tau)\geq
y_k(\tau)\geq y'' _k(\tau).
\end{equation}
Further, $x'_k(\tau )$ increases, while $y'_k(\tau )$ decreases in
$\tau$. By the same token, $x'' _k(\tau )$ decreases and $y''
_k(\tau )$ increases in $\tau$. Therefore, both pairs
$(\bfm{x}'(\tau),\bfm{y}'(\tau) )$ and $(\bfm{x}'' (\tau),\bfm{y}''
(\tau) )$ tend to limits as $\tau\to\infty$, which are fixed points,
i.e., solutions to Eqns \reff{6}. By \reff{xbod}, any solution to
(\ref{5}), (\ref{d}) eventually lies between these limits. To finish
the proof of the theorem, we have to show that the solution
$(\bfm{x}^*,\bfm{y}^*)$ to Eqn \reff{6} is unique.

For convenience, we state the corresponding assertion as Lemma
\ref{lemmauniq}.

\medskip

\begin{lemma}\label{lemmauniq} For any values $\lambda_{\RB
/\RS},\alpha, \gamma$ and $\beta\geq 0$ there exists a unique
solution to Eqns \reff{6}.
\end{lemma}

\medskip

{\it Proof of Lemma} \ref{lemmauniq}. To start with, note that every
solution to
 \reff{6} has
$$x^*_1>x^*_2>\ldots >x^*_N,\;\;y^*_1<y^*_2<\ldots <y^*_N.$$
Therefore, if $x_1\leq y_1$ then  $x_k<y_k,\ 1<k\geq N$ and if
$x_N\geq y_N$ then  $x_k>y_k,\ 1<\geq k < N$ .

 It is convenient to introduce
auxiliary variables $v_k,w_k\geq 0$ in terms of which Eqns \reff{6}
will be treated. Geometrically the idea is as follows: we start with
Eqn \reff{6}$^{(1)}$ :
 $\lambda_{\RB}=(\alpha +\beta )v_1+\min\;[v_1,w_1]$ and watch how this relation between $v_1,\
w_1$ is transformed by \reff{6}$^{(2)}$,\reff{6}$^{(3)}$ to relation
between $v_2,\ w_2$, then to relation between $v_3,\ w_3$ etc.

The locus of points $(u,v)\in\bbR^2$ where
$$v,w>0,\;\lambda_{\RB}=(\alpha +\beta )v+\min [v,w]$$
coincides with a continuous broken line, $L_1\subset\bbR^2_+$,
formed by two pieces: 1) a vertical ray $L_1^{(0)}$ emitted from the
point $o_1^{(0)}$ lying on the bisectrix where
$$o_1^{(0)}=\big(\lambda_{\RB}\big/(\alpha +\beta +\gamma),
\lambda_{\RB}\big/(\alpha +\beta +\gamma)\big)$$ and 2) a line
segment $L_1^{(1)}$ of a negative slope $-(\alpha +\beta)/\gamma$,
joining $o_1^{(0)}$ with the point $o_1^{(1)}$ on the horizontal
axis:
$$o_1^{(1)}1=\big(\lambda_{\RB}\big/(\alpha +\beta),0\big).$$
See the figure below.

\medskip
\includegraphics[width=0.8\textwidth]{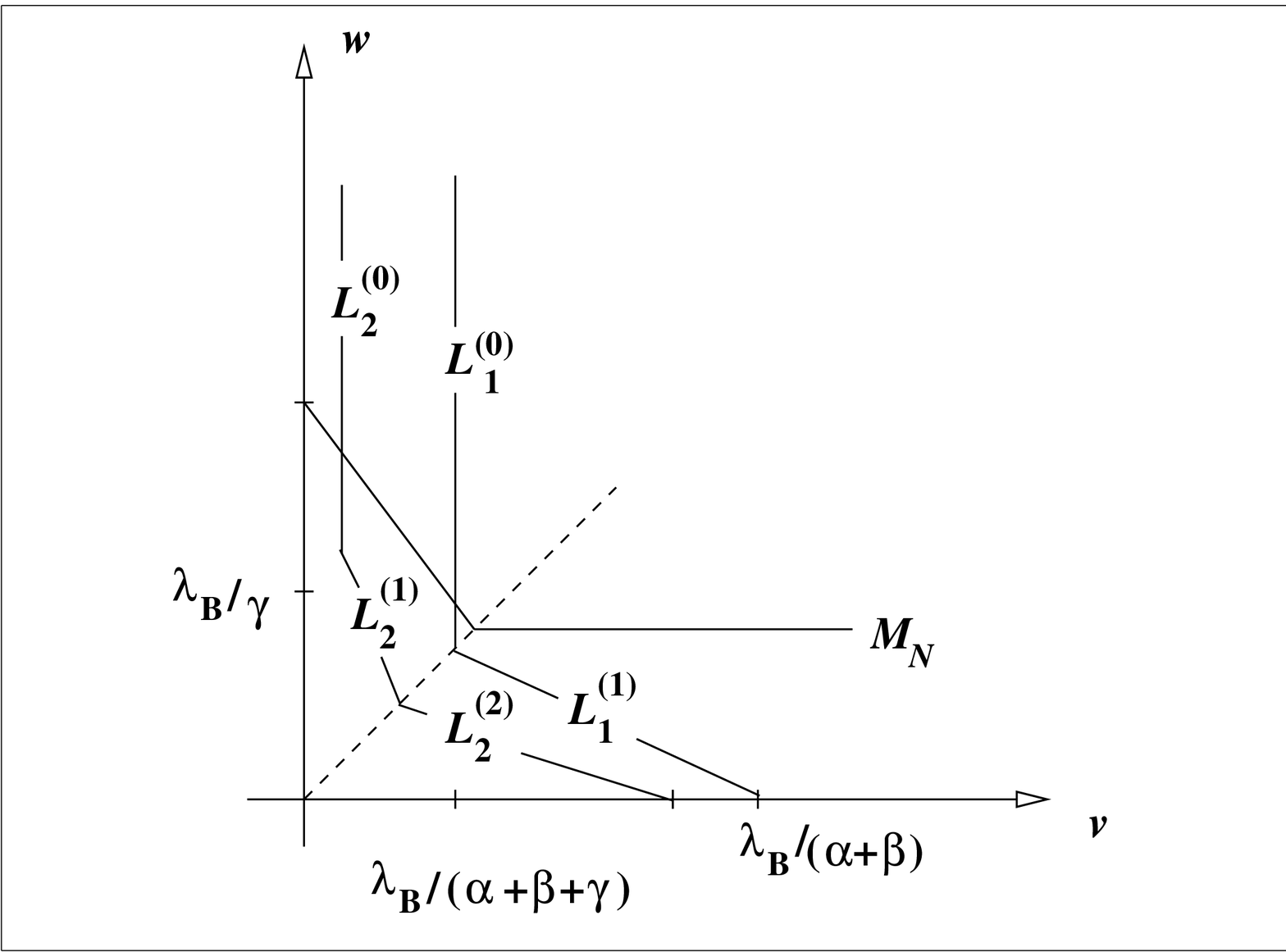}
\medskip

 The passage from values $v_1$, $w_1$ to $v_2$, $w_2$
generates a $1\ -\ 1$ map acting on $L_1$. The image of $L_1$ under
the map $(v_1,w_1)\mapsto (v_2,w_2)$ is another continuous broken
line, $L_2\subset\bbR^2_+$, formed by three pieces: 1) a vertical
ray $L_2^{(0)}$ issued from point
$$o_2^{(0)}= \big(\lambda_{\RB}\alpha\big/(\alpha +\beta +\gamma)^2,
\lambda_{\RB}\big/\alpha\big),$$ 2) a line segment $L_2^{(1)}$
joining the points $o_2^{(0)}$ and $o_2^{(1)}$ where $o_2^{(2)}$
lies on the bisectrix and has both co-ordinates equal to
$$\frac{\lambda_{\RB}\alpha (\alpha +\beta +\gamma )}{
(\alpha +\beta +\gamma )\big[(\alpha +\beta )(\alpha +\beta +\gamma
)^2+\alpha^2\gamma\big]},$$ and 3) a line segment $L_2^{(2)}$
joining $o_2^{(1)}$ with the point $o_2^{(2)}$ on the horizontal
axis:
$$o_2^{(2)}= \left(\frac{\lambda_{\RB}\alpha}{(\alpha +\beta )^2},0\right)\,.
$$
The slopes ${\rm d}w/{\rm d}v$ of segments $L_2^{(1)}$ and
$L_2^{(2)}$ are negative, but the slope flattens when we pass from
$L_2^{(1)}$ to $L_2^{(2)}$.

 In the above figure, $L_1$ and $L_2$ are shown on the same
$(v,w)$-plane $\bbR^2$; in this figure line $L_2$ lies to the left
of $L_1$. (The third broken line, $M_N$ present in the figure is
explained below.)

A similar picture persists when we iterate, i.e., pass from
$(v_2,w_2)$ to $(v_3,w_3)$ and so on. At step $k$\ \
$(v_k,w_k)\mapsto (v_{k+1},w_{k+1})$ where again the map is 1\ -\ 1,
$(v_{k+1},w_{k+1})$ belongs to a continuous broken line $L_{k+1}$ in
$\bbR^2_+$ formed by $k+2$ pieces. One piece, $L_{k+1}^{(0)}$, is a
vertical ray while the $k+1$ others, $L_{k+1}^{(1)}$, $\ldots$,
$L_{k+1}^{(k+1)}$, are line segments of negative slopes, the slope
flattens when we pass from $L_k^{(1)}$ to $L_k^{(k)}$. The last
segment, $L_{k+1}^{(k+1)}$, joins a point on the bisectrix and a
point on the horizontal axis.

 At the end of this process we obtain a continuous broken line $L_N$,
the locus of points $(v_N,w_N)$. Our next step is to consider the
intersection of $L_N$ and $M_N$ where $M_N$ is the locus where
$$\lambda_{\RS}=(\alpha +\beta )w_N+\gamma\min\;[v_N,w_N].$$
More precisely, $M_N$ is a continuous broken line formed by a
horizontal ray issued from the point
$$\left(\frac{\lambda_{\RS}}{\alpha+\beta +\gamma},
\frac{\lambda_{\RS}}{\alpha+\beta +\gamma},\right)$$ and a line
segment joining this point with the point
$$\left(0,\frac{\lambda_{\RS}}{\alpha +\beta}\right)$$
lying on the vertical axis. Cf. the figure.

We want to check that the point of intersection is always unique: it
yields to  unique  solution to \reff{6}.

 Line $L_N$ may intersect the horizontal part of $M_N$. That means
 that there exist a solution to \reff{6} where $x_1^*>y_1^*$
 and $x_N^*>y_N^*$ (this case is not presented on our figure). For the proof of
 lemma it is needed to show that
 in this case $L_N$ cannot intersect the sloppy part of $M_N$, where $y_N>x_N$.
 For that it is sufficient to prove that  ${\rm
d}w\big/{\rm d}v$ on $L_N\setminus L_N^{(N)}$, the part of $L_N$
above the bisectrix, is steeper than $-\gamma\big/(\alpha +\beta )$,
the slope of the segment of line $M_N$.

 On the other hand if $L_N$ does not intersect the horizontal part of
 $M_N$ it has to intersect the sloppy part of $M_N$ and it is needed
 to show that in this case such intersection is unique. Here again it
 is sufficient to show that
 ${\rm d}w\big/{\rm d}v$ on $L_N\setminus L_N^{(N)}$ is steeper than
$-\gamma\big/(\alpha +\beta )$.

To prove the assertion of the lemma  we show that ${\rm d}w\big/{\rm
d}v$ on $L_N\setminus L_N^{(N)}$ is always steeper than
$-\gamma\big/(\alpha +\beta )$. In fact, it suffices to verify that
\begin{equation}\label{1N}
\frac{{\rm d}w_N}{{\rm d}v_N}<-\frac{\gamma}{\alpha +\beta} \;\hbox{
on $L_N^{i},\ 1 \leq i<N $}.
\end{equation}

Any  segment $L_k^{(i)},\ 1<i<k,$  maps  onto  segment
$L_{k+1}^{(i)}$, segment  $L_k^{(k)}$ maps onto two segments:
$L_{k+1}^{(k)}$ and $L_{k+1}^{(k+1)}$. The slope of segment
$L_{k+1}^{(i)}$, $i=1,\ldots ,k+1$ is steeper than that of
$L_k^{(i)}$ and the slope of $L_{k+1}^{(k+1)}$ is steeper than that
of $L_k^{(k)}$. More, the slopes of $L_N^{(i)},\ 0<i\leq N,$ are
steeper then the steep segment of $M_N$.

To show that consider three cases of the map $L_k\to L_{k+1}$:
$$\begin{array}{l} 
1)\  v_k>w_{k},\  v_{k+1}>w_{k+1};\\ 
2)\ v_k>w_{k},\  v_{k+1}<w_{k+1};\\ 
3)\ v_k<w_{k+1},\  v_{k+1}<w_{k+1}.\end{array}$$ 
In these cases we have, the following equations
\begin{equation}\label{cases}
\displaystyle
\frac{{\rm d}w_{k+1}}{{\rm d}v_{k+1}}
=\begin{cases} \displaystyle
\frac{A_{k+1}(\alpha+\beta+\gamma)}{\alpha^2}\frac{{\rm
d}w_{k}}{{\rm d}v_{k}},\\
\displaystyle
\frac{(\alpha+\beta+\gamma)^2}{\alpha^2}
 \frac{{\rm d}w_{k}}{{\rm d}v_{k}},\\
\frac{B_{k}(\alpha+\beta+\gamma)}{\alpha^2}\frac{{\rm d}w_{k}}
 {{\rm d}v_{k}},
\end{cases}\end{equation}
where, respectively,
$$\begin{cases}
\displaystyle v_k=\frac{(\alpha+\beta )v_{k+1}+\gamma w_{k+1}}{\alpha},\
  w_{k+1}=\frac{(\alpha+\beta+\gamma)w_k}{\alpha},\\ 
\displaystyle v_k=\frac{(\alpha+\beta +\gamma )v_{k+1}}{\alpha},\
  w_{k+1}=\frac{(\alpha+\beta+\gamma)w_k}{\alpha},\\ 
\displaystyle v_k=\frac{(\alpha+\beta +\gamma)v_{k+1}}{\alpha},\
  w_{k+1}=\frac{(\alpha+\beta)w_k+\gamma v_{k}}{\alpha}.
\end{cases}$$
Here $\displaystyle A_{k+1}=\alpha+\beta +\gamma\frac{w_{k+1}}{v_{k+1}}$ 
and $\displaystyle B_{k}=\alpha+\beta +\gamma\frac{v_{k}}{w_{k}}$.

Therefore for  $N>2$
$$\begin{array}{cl}\displaystyle
\Big|\frac{{\rm d}w_{N}}{{\rm d}v_{N}}\Big|& \displaystyle > 
\Big|\frac{(\alpha+\beta+\gamma)^{N-1}}{\alpha^{N-1}}
 \frac{{\rm d}w_{1}}{{\rm d}v_{1}}\Big|=
 \frac{(\alpha+\beta+\gamma)^{N-1}}{\alpha^{N-1}}
\frac{\alpha+\beta}{\gamma}\\ \displaystyle
\displaystyle \;&\displaystyle >\Big|\frac{{\rm d}w}{{\rm
d}v}\Big|_{M_N}=\frac{\gamma}{\alpha+\beta}.\end{array}$$

For $N=2$ we have the middle case in \reff{cases} with $k=1,\ k+1=2$.
Here again the needed inequality takes place. In fact, $\frac{{\rm
d}w_1}{dv_1}=-\frac{\alpha+\beta}{\gamma}$.  By using \reff{cases},
we get that for $N=2$,
\begin{equation*}
\frac{dw_2}{du_2}=-\frac{(\alpha+\beta+\gamma)^2}{\alpha^2}\frac{dw_1}{du_2}
 \;\hbox{
on $L_2^{1}$}.
\end{equation*}
This finishes the proof of Lemma \ref{lemmauniq}. $\blacktriangle$

\medskip

\begin{theorem}\label{thm3}  Suppose that the re-scaled initial states
converge in probability:
for any $\epsilon >0$,
\begin{equation*}
\lim_{L\to\infty}\bbP\left(\dist\left[{\diy\frac{1}{L}}\bfm{U}(0),
(\bfm{x}(0),\bfm{y}(0))\right]\,\geq\epsilon\right) =0.
\end{equation*}
Then, for all $T>0$, the process $\Big\{\bfm{V}^{(L)}(\tau ),\;
\tau\in [0,T]\Big\}$ converges in probability to the solution
$\Big\{(\bfm{x}(\tau ),\bfm{y}(\tau )),\;0\leq\tau\leq T\Big\}$.
That is, $\forall$ $\epsilon >0$,
\begin{equation}\label{9}
\lim_{L\to\infty}\bbP\left(\operatornamewithlimits{\sup}\limits_{0\leq\tau
\leq T} \left\{\dist\left[\bfm{V}^{(L)}(\tau ),
(\bfm{x}(\tau),\bfm{y}(\tau))\right]\right\} \geq\epsilon\right)=0.
\end{equation}

In particular, if $\bfm{x}(0)=\bfm{x}^*$ and
$\bfm{y}(0)=\bfm{y}^*$ then
\begin{equation}\label{10}
\lim_{L\to\infty}\bbP\left(\operatornamewithlimits{\sup}\limits_{0\leq\tau
\leq T} \left\{\dist\left[\bfm{V}^{(L)}(\tau ),
\big(\bfm{x}^*,\bfm{y}^*\big)\right]\right\} \geq\epsilon\right)=0.
\end{equation}
 Moreover, if process $\big\{\bfm{U}(t),\big\}$ is in equilibrium
then Eqn \reff{10}  holds true.
\end{theorem}

\medskip

{\it Proof of Theorem} \ref{thm3}. Let $\bfm{G}^{(L)}$ denote the
generator of the Markov process $\{\bfm{U}^{(L)}(t)\}$ (with rates
as in Eqn (\ref{3})). Then the action of matrix $\bfm{G}^{(L)}$ on
functions $\phi (\bfm{b},\bfm{s})$ of state variables
$\bfm{b},\bfm{s}\in\bbZ_+^N$ is determined by the equation
\begin{equation}\label{8A}\begin{array}{l}\bfm{G}^{(L)}\phi (\bfm{b},\bfm{s})= \diy\frac{\beta }{L}\sum\limits_{1\leq k\leq N}\Big(
\big[\phi (\bfm{b}-\bfm{e}_k,\bfm{s})-\phi (\bfm{b},\bfm{s})\big]b_k\\
\qquad\qquad\qquad\qquad\qquad\qquad +\big[\phi
(\bfm{b},\bfm{s}-\bfm{e}_k)-\phi (\bfm{b},\bfm{s})\big]s_k
\Big)\\
\qquad\qquad\quad +\diy\frac{\alpha }{L}\bigg(\sum\limits_{1\leq k< N}\big[\phi (\bfm{b}-\bfm{e}_k+\bfm{e}_{k+1},\bfm{s})-\phi (\bfm{b},\bfm{s})\big]b_k\\
\qquad\qquad\qquad\qquad\diy +\sum\limits_{1< k\leq N}\big[\phi (\bfm{b},\bfm{s}-\bfm{e}_k+\bfm{e}_{k-1})-\phi (\bfm{b},\bfm{s})\big]s_k\bigg)\\ \;\\
\qquad\qquad\quad +\diy\lambda_{\RB}\big[\phi
(\bfm{b}+\bfm{e}_1,\bfm{s})-\phi (\bfm{b},\bfm{s})\big]
+\diy\lambda_{\RS}\big[\phi (\bfm{b},\bfm{s}+\bfm{e}_N)-\phi (\bfm{b},\bfm{s})\big]\\
\;\\
\qquad\qquad\quad +\diy\frac{\gamma}{L}\sum\limits_{1\leq k\leq N}
\big[\phi (\bfm{b}-\bfm{e}_k,\bfm{s}-\bfm{e}_k)-\phi
(\bfm{b},\bfm{s})\big]\big(b_k\wedge s_k\big)\,.
\end{array}\end{equation}
Here $\bfm{e}_k$, $1\leq k\leq N$, stands for the vector in
$\bbZ_+^N$ whose components are all $0$'s except for the $k$th one,
equal to $1$.

In particular, we can take a function $\phi=\phi^{(L)}$ of the
form $\phi (\bfm{b},\bfm{s})=\varphi\left(\diy\frac{\bfm{b}}{L},
\frac{\bfm{s}}{L}\right)$ where $\varphi$ is a smooth function on
$\bbR_+^N$. This choice agrees with the spatial scaling $\bfm{x}\sim\bfm{b}\big/L$,
 $\bfm{y}\sim\bfm{s}\big/L$ in Eqn (\ref{3A}). Then $\forall$ $\bfm{x}=(x_1,\ldots ,x_N)\in\bbR_+^N$
and $\bfm{y} =(y_1,\ldots ,y_N)\in\bbR_+^N$, with $\bfm{b}=\lfloor L\bfm{x}\rfloor$,
$\bfm{s}=\lfloor L\bfm{y}\rfloor$ where $\lfloor\,\cdot\,\rfloor$ stands for
the integer part, we obtain that
$$\begin{array}{l}
\bfm{G}^{(L)}\varphi\big(\lfloor L\bfm{x}\rfloor\big/L,
\lfloor L\bfm{y}\rfloor\big/L\big)\\
\qquad =\diy\beta \sum\limits_{1\leq k\leq N}\bigg\{
\Big[\varphi\Big(\big(\bfm{b}-\bfm{e}_k\big)\big/L,\bfm{s}\big/L\Big)
-\varphi\Big(\bfm{b}\big/L,\bfm{s}\big/L\Big)\Big]b_k\big/L\\
\qquad\qquad\qquad +\Big[\varphi
\Big(\bfm{b}\big/L,\big(\bfm{s}-\bfm{e}_k\big)\big/L\Big)-\varphi \Big(\bfm{b}\big/L,\bfm{s}\big/L\Big)\Big]s_k\big/L
\bigg\}\\
\qquad +\diy\alpha\bigg\{\sum\limits_{1\leq k< N}\Big[\varphi \Big(\big( \bfm{b}-\bfm{e}_k+\bfm{e}_{k+1}\big)\big/L,\bfm{s}\big/L\Big)
-\varphi\Big(\bfm{b}\big/L,\bfm{s}\big/L\Big)\Big]b_k\big/L\\
\quad\qquad\diy +\sum\limits_{1< k\leq N}\Big[\varphi\Big( \bfm{b}\big/L,\big(\bfm{s}-\bfm{e}_k+\bfm{e}_{k-1}\big)\big/L\Big)-\varphi\Big( \bfm{b}\big/L,\bfm{s}\big/L\Big)\Big]s_k\big/L\bigg\}\\ \;\\
\qquad\qquad\qquad\quad +\diy\lambda_{\RB}\Big[\varphi
\Big(\big(\bfm{b}+\bfm{e}_1\big)\big/L,\bfm{s}\big/L\Big)-\varphi
\Big(\bfm{b}\big/L,\bfm{s}\big/L\Big)\Big]\\
\qquad\qquad\qquad\quad +\diy\lambda_{\RS}\Big[\varphi\Big(\bfm{b}\big/L,\big(\bfm{s}
+\bfm{e}_N\big)\big/L\Big)-\varphi\Big(\bfm{b}\big/L,\bfm{s}\big/L\Big)\Big]\\
\;\\
\quad +\diy\gamma\sum\limits_{1\leq k\leq N}
\Big[\varphi\Big(\big(\bfm{b}-\bfm{e}_k\big)\big/L,\big(\bfm{s}-
\bfm{e}_k\big)\big/L\Big)-\varphi\Big(
\bfm{b}\big/L,\bfm{s}\big/L\Big)\Big]\big(b_k\wedge s_k\big)\big/L\,.
\end{array}$$
Next, we multiply the both side by $L$ -- in agreement with the
time-scale $\tau\sim t\big/L$ in Eqn (\ref{3A}) -- and pass to the
limit $L\to\infty$. This yields
$$\begin{array}{l}
\diy\lim_{L\to\infty}L\bfm{G}^{(L)}\varphi (\bfm{x},\bfm{y})\\
\qquad\diy = \Bigg\{-\beta \sum\limits_{1\leq k\leq
N}\Big(x_k\frac{\partial}{\partial x_k}
+y_k\frac{\partial}{\partial y_k}\Big)\\
\qquad\diy +\alpha \Big[\sum\limits_{1\leq k<
N}x_k\Big(\frac{\partial}{\partial x_{k+1}}-\frac{\partial}{\partial
x_{k}}\Big) +\sum\limits_{1< k\leq N}
y_k\Big(\frac{\partial}{\partial y_{k-1}}-
\frac{\partial}{\partial y_{k}}\Big)\Big]\\
\quad\diy +\left(\lambda_{\RB}\frac{\partial}{\partial x_{1}}
+\lambda_{\RS}\frac{\partial}{\partial y_{N}}\right)
-\gamma\sum\limits_{1\leq k\leq N}\big(x_k\wedge
y_k\big)\Big(\frac{\partial}{\partial x_i}
+\frac{\partial}{\partial y_i}\Big)\Bigg\}\varphi (\bfm{x},\bfm{y})\,.\\
\;\end{array}$$ Applying Theorem 6.1 from \cite{EK}, we obtain Eqn
(\ref{9}).\def\wh{\widehat}

The next remark is that each scaled process $\{\bfm{V}^{(L)}\}$ has
a unique invariant distribution $\wh{\pi}^{(L)}$; the family of
probability distributions $\wh{\pi}^{(L)}$ (considered on
$\bbR_+^N$) is compact in the sense of convergence in probability.
This can be deduced from the above remark that the original
processes $\{\bfm{U}^{(L)}\}$, and hence, the scaled process
$\{\bfm{V}^{(L)}\}$ can be majorized by suitable analogs of
M/M/$\infty$ queueing systems. It is easy to see that every limiting
point for $\wh{\pi}^{(L)}$ when $L\to\infty$ is a delta-measure
sitting at a fixed point for system (\ref{d}). However, the latter
is unique and coincides with $(\bfm{x}^*,\bfm{y}^*)$. Consequently,
the distributions $\wh{\pi}^{(L)}$ converge in probability to the
aforementioned delta-measure. Then, applying the already established
result, we obtain Eqn (\ref{10}).
$\quad\blacktriangle$

\section{ Fixed points in the scaling limit. Concluding
remarks}

The approximation developed in Theorem 3 calls for an analysis of
solutions to \reff{6}.

The parameter space $\bbR^5_+$ formed by $\gamma$, $\alpha_{{\rm
Q}/{\rm M}}$, and $\lambda_{{\rm b}/{\rm s}}$ is partitioned into
open domains where one of the following generic patterns persists:

(i) $v_N>w_N$, (ii) $v_1<w_1$,  and
 (iii) $v_i>w_i$
for $i=1,\ldots,\ell$ and $v_i<w_i$ for $i=\ell +1,\ldots ,N$
where $1<\ell<N$. In each of these domains system \reff{6}  is
linear.

\medskip

A particular algorithm for calculating
$\big(\bfm{x}^*,\bfm{y}^*\big)$ is based on the following recursion.
Set
\begin{equation*}
x^{(0)}_1=\frac{\lambda_b}{\alpha_{\rm Q}+ \alpha_{\rm M}+\gamma}, \
 x^{(0)}_i=\frac{\alpha_{\rm M}x^{(0)}_{i-1}}{\alpha_{\rm Q}+
\alpha_{\rm M}+\gamma},\
y^{(0)}_N=\frac{\lambda_s}{\alpha_{\rm Q}+ \alpha_{\rm M}}, \
 y^{(0)}_i=\frac{\alpha_{\rm M}y^{(0)}_{i+1}}{\alpha_{\rm Q}+
\alpha_{\rm M}}
\end{equation*}
  Next, let
$\big(\bfm{x}^{(k)},\bfm{y}^{(k)}\big)$, $k=1,2,\dots$ be the
solution to the system
\begin{equation*}
\begin{array}{ll}
\lambda_{\rm b}&=\Big(\alpha_{q}+\alpha_{m}\Big)x^{(k)}_1
 +\gamma\min\;\big[x^{(k-1)}_1,y^{(k-1)}_1\big],\\
\alpha_{m}x^{(k)}_{i-1}&=\Big(\alpha_{q}+ \alpha_{m}\Big)x^{(k)}_i
 +\gamma\min\;\big[x^{(k-1)}_i,y^{(k-1)}_i\big],\;1<i\leq N,\\
\alpha_{m}y^{(k)}_{i+1}&=\,\Big(\alpha_{q}+ \alpha_{m}\Big)y^{(k)}_i
 +\;\gamma\min\;\big[x^{(k)}_i,y^{(k-1)}_i\big],\;1\leq i< N,\\
\lambda_{\rm s}&=\Big(\alpha_{q}+\alpha_{m}\Big)y^{(k)}_N
+\gamma\min\;\big[x^{(k)}_N,y^{(k-1)}_N\big].\end{array}
\end{equation*}

 These iterations converge because the inequalities $x^{(k)}_i\ >\ x^{(k-1)}_i$,\ $y^{(k)}_i\ <\
y^{(k-1)}_i$\ hold true $\forall \ i,k\geq 1$\ and,  values
$x^{(k)}_i,y^{(k)}_i $ are uniformly bounded and there exist
$\lim\limits_{k\to \infty}x^{(k)}_i$,$\lim\limits_{k\to
\infty}y^{(k)}_i$ that, naturally,  satisfy \reff{6}.

\medskip

We conclude with the following remarks.

\medskip

Our model presents also a caricature of "overproduction crisis":
In fact, if $\lambda_s$ is sufficiently large, so that $x_i<\y_i,\
1\leq i\leq N,$ then  the amount of "trades" $\sum_i\gamma \min
[x_i,y_i] = \sum_i\gamma x_i$ is not changing  by increase of
$\lambda_s$, all "extra" sellers leave the market without performing
any trade.

It is interesting to investigate the dependance of trade performance
on $\gamma$ as $\gamma \to \infty$, that is where the trade action
happens almost immediately after the moment when traders appear at
some price level.   Then almost all trades happen at two levels
$i_0$ and $i_0+1$, and $x_i$ is very small as $i>i_0+1$, $y_i$ ia
very small as $i<i_0$. But our limit model  does not permit to
consider the case $\gamma=\infty$, though, sure the initial Markov
process can be investigated in case of immediate trade deals.
The limiting case $\gamma=\infty$ of our model is close to the problems
investigated in  \cite{MMZ}.

\medskip

We hope that the variation of these models parameters can help to
determine factors attracting or repelling various `market
participants'. An important aspect of any model of the market is
what possibilities it gives for an accurate  prediction of the
stochastic component in the dynamics of the market prices and
volumes.

The current set-up of the model presented here admits
straightforward generalizations to the case where parameters
$\gamma$ and $\alpha_{Q/M}$ depend on $i$, $0<i<N$ and on the trader
type (b/s). Another generalization emerges if these parameters and
$\lambda_{b/s}$ become state-dependent. It is also possible to allow
the exogenous buyers and sellers to enter the system at any price
level among $c_1$, $\ldots$, $c_N$. To take into account elements of
the FCFS discipline, one could introduce various priorities into the
dynamics of process $\bfm{U}(t)$.

Finally, we would like to note that there are several forms of
convergence for which the
assertion in Theorem 3 holds true.

\section*{Acknowledgements}
The authors thank the anonymous referees
for stimulating critical remarks. NV would like to thank the  grant
RFBR 11-01-00485-a. YS would like to thank the FAPESP Foundation,
Sao Paulo, Brazil, for providing a grant towards this work,
and NUMEC/IME, University of Sao Paulo, for warm hospitality.

\medskip


N.Vvedenskaya, Institute for Information
Transmission\\ Problems, RAS, GSP-4, Moscow 127994, RUSSIA; {\tt
ndv$@$iitp.ru }

Y.Suhov, Institute for Information
Transmission Problems, RAS, GSP-4, Moscow 127994, RUSSIA; 
DPMMS, University of Cambridge, and St John's College, Cambridge 
CB3 0WB, UK; IME,
Universidade de Sao Paulo, C.P. 66281 CEP 05389-970 Sao Paulo,
BRAZIL; {\tt yms@statslab.cam.ac.uk}

V.Belitsky, IME,
Universidade de Sao Paulo, C.P. 66281 CEP 05389-970 Sao Paulo,
BRAZIL; {\tt  belitsky@ime.usp.br}

\end{document}